# Analysis of skin tissues spatial fluorescence distribution by the Monte Carlo simulation

D Y Churmakov[1], I V Meglinski[1], S A Piletsky[2] and
D A Greenhalgh[1]

[1] School of Engineering, Cranfield University, Cranfield, MK43 0AL, UK
[2] Institute of BioScience and Technology, Cranfield University, Silsoe, MK45 4DT, UK

**Abstract**
A novel Monte Carlo technique of simulation of spatial fluorescence distribution within the human skin is presented. The computational model of skin takes into account the spatial distribution of fluorophores, which would arise due to the structure of collagen fibres, compared to the epidermis and stratum corneum where the distribution of fluorophores is assumed to be homogeneous. The results of simulation suggest that distribution of auto-fluorescence is significantly suppressed in the near-infrared spectral region, whereas the spatial distribution of fluorescence sources within a sensor layer embedded in the epidermis is localized at an 'effective' depth.

## 1. Introduction

Optical techniques have recently been received considerable attention in the fields of biomedical diagnostics and monitoring of biological tissues [1, 2]. The fluorescence spectroscopy is notable among other non-invasive diagnostic techniques, as it offers an exciting precision, selectivity and sensitivity to the biochemical make-up of tissues [3–5].

Recently, a new non-invasive optical/fluorescence technique for express clinical diagnostics and therapeutic monitoring of skin has been proposed [6]. This new technique is based on scanning a 'tattoo' pattern transferred on the skin in a similar manner to children's non-permanent 'tattoos'. Typically, the later are simply removed by several washings. By incorporating 'smart' polymer nano-particles within the 'tattoo' formulation it will be possible to generate fluorescence signals with specific spectral signatures that are indicative of the state of the tissue. Thus, physiological changes pertaining to temperature, the concentration of metabolites or the presence of drugs can be determined. Since a 'tattoo' can combine several chemical sensors, either by intermixing or by separate patterning on to the skin, multi-parameter measurements are possible. This new methodology has considerable clinical potential both in hospitals and in surgeries. We envisage usage for routine monitoring as well as for the more complex therapeutic management of drug administration. Other important potential applications include: early warning of excessive exposure to ultraviolet radiation, general health monitoring, fundamental physiological investigations, measuring sensitivities to cosmetics or household products, allergy detection, etc.

The total fluorescence of skin tissues-'tattoo' comprises of the auto-fluorescence, i.e. the fluorescence of endogenous fluorophores such as amino acids (tryptophan, tyrosine) and structural proteins (collagen and elastin) randomly distributed within the skin, and the fluorescence of the exogenous fluorophores associated with the 'tattoo' pattern. The turbidity of skin tissues hampers the interpretation of the direct fluorescence measurements as intrinsic fluorescence [7, 8]. The analysis of the contribution of endogenous and exogenous fluorophores to the detected signal requires a knowledge of the spatial distribution of the fluorescence sources within the tissues. We have investigated where the fluorescence is excited in tissues and how the distribution of fluorescence sources depends on the fluorophore parameters as well as optical variations in the tissue. The goal of this study is to estimate the effect of fluorescence sources heterogeneously distributed in a model of skin. To predict the fluorescence distribution in biological media, various techniques have been developed and used. These include electromagnetic theory [9], Kubelka–Munk approximation [10], the diffusion theory [11, 12], random walk theory [13–15] and Monte Carlo (MC) techniques [16–23]. The MC technique has a number of advantages over analytical models: different boundary conditions can be accounted for; the technique allows the investigation of various phenomena; the method is suitable for both highly scattering and absorbing multilayered media; the technique may also be adapted for fluorescence modelling. This report describes a novel MC technique used to simulate spatial distribution of fluorescence excitation in a model of human skin.

## 2. Method of simulation

### 2.1. MC simulation of light propagation in a medium

The stochastic numerical MC method is widely used to model optical radiation propagation in complex randomly inhomogeneous highly scattering and absorbing media such as biological tissues [24–30]. Basic MC modelling of an individual photon packets trajectory consists of the sequence of the elementary simulations [24–28]: photon pathlength generation, scattering and absorption events, reflection or/and refraction on the medium boundaries. The initial and final states of the photons are entirely determined by the source-detector geometry. The photons packets are launched in the medium within a uniform random distribution over the angles defined by numerical aperture of source. At the scattering site a new photon packet direction is determined according to the Henyey–Greenstein scattering phase function [31].

In contrast to previous works [24–28], we use an MC technique, which combines the statistical weight scheme and effective optical photon paths [29, 30]. In framework of this approach, absorption occurs between the scattering events only. In other words, the medium consists of a scattering-centre matrix embedded within an absorbing continuum that is consistent with the microscopic Beer–Lambert law. Absorption is realized by recalculating the statistical weight of each photon packet according to its pathlength between source and detector areas [29, 30]:

$$W = W_0 \exp\left(-\sum_{i=1}^{N} \mu_a l_i\right). \quad (1)$$

Here, $W$ is the statistical weight of a photon packet at the $N$th step of propagation within the medium, $W_0$ is the initial weight of the photon packet, $\mu_a$ is the absorption coefficient per unit pathlength $l_i$, $N$ is the total number of scattering events undergone by the photon packet during its random walk. The internal photon packet reflection/refraction on the medium boundary are taken into account by splitting the photon packet into the reflected and the transmitted parts [30, 32]. This procedure is important for the shallow probing of skin tissues [29, 32]. The simulation of a photon packet tracing is truncated when its statistical weight is less than $10^{-4}$, or a photon packet has been scattered more than $10^4$ times. Beyond these limits the statistical weight of a photon packet is too small and the packet no longer contributes to the fluorescence excitation, or the packet has propagated too far from the area of interest. Typically $10^5$–$10^7$ packets of detected photons are simulated.

The above description of the MC technique has been validated against analytic solution of the photon diffusion equation for semi-infinite homogeneous scattering medium [30]. The result of the MC simulation of the radial intensity distribution compared to the improved diffusion theory demonstrates a good agreement (figure 1). It also demonstrates that when a computational model of skin is used with reasonable physical and structural parameters, the results of skin diffuse reflectance spectra simulation agree reasonably well with the results of *in vivo* skin spectra measurements [33] (figure 2).

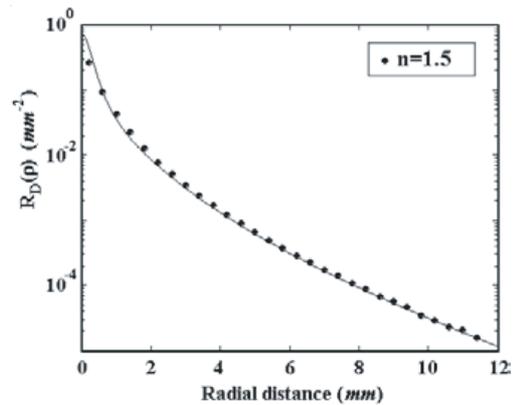

**Figure 1.** The results of MC model validation: radial distribution of the intensity of diffusely reflected radiation on the medium surface predicted by an improved diffusion theory (——) and by MC simulation (♦) [30]. The optical parameters used in the simulation are: $\mu_s = 30$ mm$^{-1}$, $\mu_a = 0.01$ mm$^{-1}$ and $g = 0.9$, $n = 1.5$.

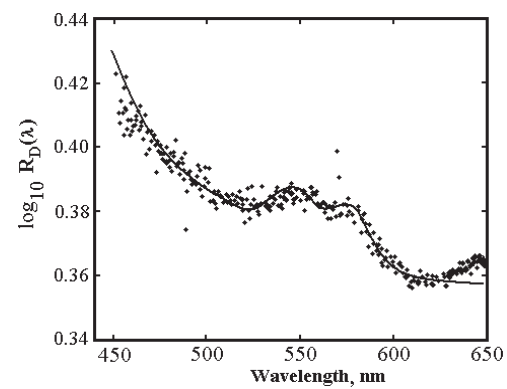

**Figure 2.** The results of MC model validation: the reflectance spectra of skin: dots, the measured *inv vivo*; ——, the results of MC simulation [33].

### 2.2. Fluorescence simulation

Earlier MC schemes of fluorescence modelling consist of the three simulation steps [16, 17, 19, 20]. First, the fluence rate distribution within a tissue volume is calculated by the standard MC scheme [24–28]. At the second step, spatial fluorescence distribution is obtained by multiplying the fluence rate distribution to the intrinsic fluorescence profile, which is defined as the product of the absorption coefficient of the fluorophore at the excitation wavelength and its quantum yield at the emission wavelength [19]. Finally, the detected fluorescence is calculated as the convolution of the fluorescence source distribution throughout the tissue with a Green function [16, 17, 19, 20]. In the framework of this model, the intensity of the simulated local fluorescence is proportional to the fraction of the absorbed energy that is determined by quantum yield of the fluorophore. The fluorescence source distribution within the medium is mainly dependent on the fluence rate distribution. Crilly *et al* [18] employed the MC fluorescence forward-adjoin model. This MC scheme utilizes the solution of a transport equation both in forward (excitation photon) and in adjoin (fluorescent photon) calculations. The solution of the adjoin transport equation is obtained for those fluorescence photons that contribute to the detected fluorescence signal.

More recently, another scheme of an independent simulation of the fluorescence acts has been proposed [21]. In this later approach, the emission of the fluorescence photons occurs at the scattering sites and the quantum yield of a fluorophore $\gamma$ serves as the fluorescence threshold probability (figure 3(*a*)). The intensity of simulated fluorescence is defined by the fraction of the absorbed radiation $W_{i-1} - W_i$ (see figure 3(*a*)). In a more plausible model of fluorescence simulation [22, 23], the fluorophore absorption $\mu_a^f$ is separated from the total medium/layer absorption by the standard rejection scheme based on the fluorophore absorption threshold $P_a = (1 - \exp(-\mu_a^f l_i))$ (figure 3(*b*)). Here, the intensity of generated fluorescence is equal to the product of the quantum yield and the intensity of the incident radiation $\gamma W_{i-1}$. In this model, each photon packet produces only one fluorescence event. Both models assumed that the fluorescence is emitted uniformly from the scattering sites in random directions (see figures 3(*a*) and (*b*)).

Below we present an extension of the MC technique, [29, 30] which has been described briefly in section 2.1, for fluorescence simulation. The schematic of the fluorescence simulation is given in figure 4. The probability of the fluorescence excitation is determined as:

$$Q(x, y, z) = W P_a P_\rho P_\gamma, \quad (2)$$

where $W$, defined by equation (1), is the probability that photon packet has reached a point $(x, y, z)$ in the medium; $P_a$ is the probability of the photon packet absorption; $P_\rho$ is the probability of absorption by the fluorophore non-uniformly distributed within the medium; $P_\gamma$ is the probability of the fluorescence determined by the fluorophore quantum yield $\gamma$. The probabilities $P_a$, $P_\rho$, $P_\gamma$ are calculated by the standard rejection scheme [34]. In contrast to the above-mentioned models (see figures 3(*a*) and (*b*)), where fluorescence is emitted at the scattering sites, we define the origin of florescence at an arbitrary point $(x, y, z)$ between the scattering events, i.e. solely in the absorption sites (see figure 4). However, distribution of the fluorophores within the human skin is complex [1, 7, 35]. The stratum corneum and the epidermis mainly contain NAD(H), elastin, keratin, flavins and some other fluorophores randomly distributed within these layers [1, 35, 36]. In the dermal layers the spatial distribution of the fluorophore closely follows the distribution of collagen fibres. The latter are organized in long, wavy bundles, which vary in diameter between 1 and 40 $\mu$m [37–39]. Collagen bundles interweave in a complex and random manner to form a three-dimensional irregular meshwork (figure 5). We describe this meshwork as:

$$\rho(\mathbf{r}, z) = \cos(k\mathbf{r}) \cos(kz), \quad (3)$$

where $k = \pi/d$, $d$ is the collagen fibre diameter, $\mathbf{r} = \mathbf{r}(x, y)$ and $z$ are the coordinates of a point in the medium. This non-homogeneous distribution of fluorescence within dermal layers is clearly illustrated in the experimental images of auto-fluorescence of human skin (figure 6), whereas the distribution of fluorophores in stratum corneum and epidermis appears to be homogeneous [19, 35]. The current MC model neglects all polarization effects that might result in anisotropic fluorescence emission [7]. Therefore, the fluorescence photons are emitted isotropically from the source points, this agrees with the assumptions proposed in [16–23].

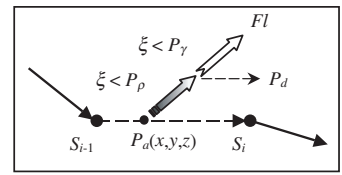

**Figure 4.** Schematic representation of the fluorescence modelling: $P_a$ is probability of the photon packet absorption between two scattering sites $S_{i-1}$ and $S_i$; $P_\rho$ is the probability of the absorption by the fluorophore; $P_\gamma$ is the probability of the fluorophore fluorescence determined by the fluorophore quantum yield $\gamma$; $P_d = (1 - P_\gamma)$ is the probability of dissipation determining a fraction of absorbed energy exerted non-radiative relaxation through other mechanisms, e.g. thermal excitation, phosphorescence etc; $\rho(x, y, z)$ determines spatial fluorophores distribution within the medium.

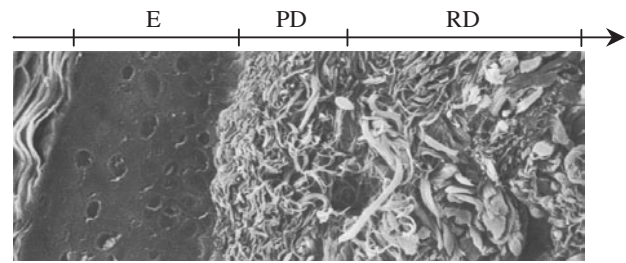

**Figure 5.** The scanning electron micrograph taken from [37] shows the arrangement of collagen fibres in the dermis (with permission). Fibre bundle diameters and density of packing in the papillary dermis (PD) and reticular dermis (RD) are different. Collagen is organized in long, wavy bundles, which vary in diameter from about 1 to 40 $\mu$m. Collagen bundles interweave in a complex and random manner to form a three-dimensional irregular meshwork [38, 39].

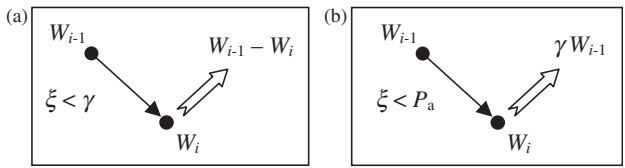

**Figure 3.** Schematic representation of the fluorescence simulation: (*a*) the fluorescence probability is determined by quantum yield $\gamma$ of a fluorophore [21]; (*b*) each fluorescence event is determined by the probability of the photon packet absorption $P_a = (1 - \exp(-\mu_{af} l_i))$ [22, 23]. Here, $W_{i-1}$ and $W_i$ are the statistical photon weights at the $(i - 1)$th and $i$th steps of photon packet, respectively; $\mu_a^f$ is the fluorophore absorption coefficient; $l_i$ is the pathlength of a photon between the scattering events; $\xi$ ($0 \leqslant \xi \leqslant 1$) is the uniformly distributed random number used in the simulation.

## 3. Results and discussion

The described MC algorithm has been implemented for the prediction of spatial distribution of the skin auto-fluorescence and 'tattoo' pattern fluorescence. Optical properties of skin tissues were estimated according to [8, 33] (see table 1 for the details). Parameters of the sensor layer are assumed to be close to optical characteristics of the most dominant fluorophores used in diagnostic measurements [40]. The 488 nm wavelength was chosen as an excitation wavelength, since it is close to the fluorescein absorption maximum (494 nm) [40]. The diameters of collagen bundles were chosen as 3, 6, 20, 30 $\mu$m in the papillary dermis (PD), upper blood net dermis, dermis, deep blood net dermis, respectively (see table 1).

The results of simulation of spatial distribution of the auto-fluorescence excitation probabilities within human skin are shown in figure 7. The observed 'porous', periodical structure of the auto-fluorescence sources distribution in the skin (see figure 7(a)) is the consequence of the collagen meshwork simulated by equation (3). The spatial distribution of the fluorescence excitation within dermal layers has a distinct periodical structure, both seen in figures 7(a) and (b). The period of this structure is close to the collagen bundle diameter of modelled layers, 20 $\mu$m (see table 1). The PD lies at 150–250 $\mu$m depth and contains small (0.3–3 $\mu$m in diameter [38, 39]), loosely distributed collagen fibres (see figure 5). Consequently, the distribution of fluorescence sources in the PD seems highly granular, and the fine porous structure is marginally observed (see figure 7(a)). Figure 7(b) gives a profile perception of the fluorescence excitation distribution along the axis $z$. This illustrates the texture profile of the fluorescence excitation within the reticular dermis (RD) resulting from large diameter of collagen fibres (10–40 $\mu$m on average [38, 39]). In comparison, in the uppermost part of dermal layers (250–300 $\mu$m) with collagen bundle diameter of 6 $\mu$m, the fluorescence sources structure is not obviously distinguished (see figure 7(b)). The calculated probability of auto-fluorescence excitation in dermal layers is significantly higher than in stratum corneum and epidermis. This agrees well with experimental data [19, 35].

The new MC model was also employed to simulate the fluorescence of the exogenous ('tattoo') fluorophores. The 'tattoo' pattern was modelled as a plane sensor layer, 50 $\mu$m thick embedded within the epidermis (100–150 $\mu$m). The thickness of pure epidermis in this simulation was chosen as 80 $\mu$m (see table 1). The result of simulation shows a similar porous structure of the auto-fluorescence sources distribution within the dermal layers (figure 8(a)). But the results also predict that the fluorescence excitation in sensor layer (100–150 $\mu$m) and in dermal layers (250–330 $\mu$m) is comparable (see figure 8(b)).

The auto-fluorescence is greatly reduced provided the tissue is illuminated at longer excitation wavelength [1, 7]. Additional MC simulations were carried out to illustrate how localization of the fluorescence sources excitation is affected by increasing the excitation wavelength towards the near-infrared. The modelled skin tissues optical properties were chosen for 700 nm (see table 1). The absorption coefficients $\mu_a$ of skin layers are taken to be a factor of ten less [8], however, due to the monotonically decreasing scattering of skin tissue with wavelength in the range 450–1100 nm [41–43] the scattering coefficients $\mu_s$ are reduced by a factor of 2–3 (see table 1). The rest of the optical properties and fluorescence parameters are assumed to be independent of wavelength, except for the optical parameters of the sensor layer.

The results of the spatial distribution of the fluorescence sources are shown in figure 9(a). The auto-fluorescence excitation of dermal layers is now highly suppressed due to the low fluorescence efficiency of natural fluorophores in the near-infrared spectral region [7]. However, it is still observed in deeper skin layers due to large penetration depth of the optical radiation in the near-infrared spectral region [8] (see figure 9), whereas the main fluorescence excitation is localized in the sensor layer (see figure 9(b)).

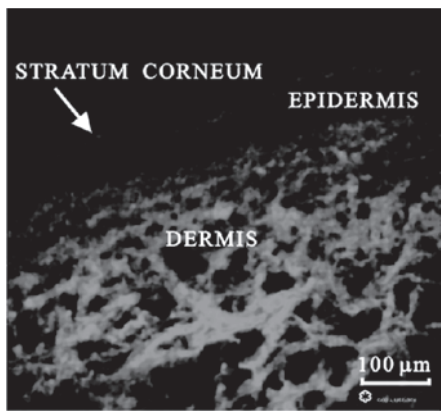

**Figure 6.** Experimental auto-fluorescence image of skin tissue section under illumination of 442 nm laser radiation (courtesy of Zeng [19], with permission).

**Table 1.** Optical properties of computational model of skin.

| k | Skin layer | $\mu_s$ (mm$^{-1}$) | | $\mu_a$ (mm$^{-1}$) | g | n | t ($\mu$m) | d ($\mu$m) | $\gamma$ |
| | | 488 nm | 700 nm | | | | | | |
|---|---|---|---|---|---|---|---|---|---|
| 1 | Stratum corneum | 40 | 20 | 0.2 | 0.9 | 1.5 | 20 | — | 0.01 |
| 2 | Epidermis | 35 | 10 | 0.15 | 0.85 | 1.34 | 130/80 | — | 0.01 |
| 3 | Sensor layer | 5 | 5 | 0.1 | 0.6 | 1.37 | 50 | — | 0.7 |
| 4 | Papillary dermis | 30 | 12 | 0.7 | 0.8 | 1.4 | 100 | 3 | 0.15 |
| 5 | Upper blood net dermis | 35 | 15 | 1.0 | 0.9 | 1.39 | 80 | 6 | 0.15 |
| 6 | Dermis | 27 | 12 | 0.7 | 0.76 | 1.4 | 1500 | 20 | 0.15 |
| 7 | Deep blood net dermis | 35 | 15 | 1.0 | 0.95 | 1.39 | 200 | 30 | 0.15 |
| 8 | Subcutaneous fat | 15 | 5 | 0.3 | 0.8 | 1.44 | 5000 | — | 0.001 |

$\mu_s$—scattering coefficient, $\mu_a$—absorption coefficient, g—anisotropy factor, n—refractive index, t—thickness of a layer, d—diameter of a collagen fibre, $\gamma$—fluorescence quantum yield.

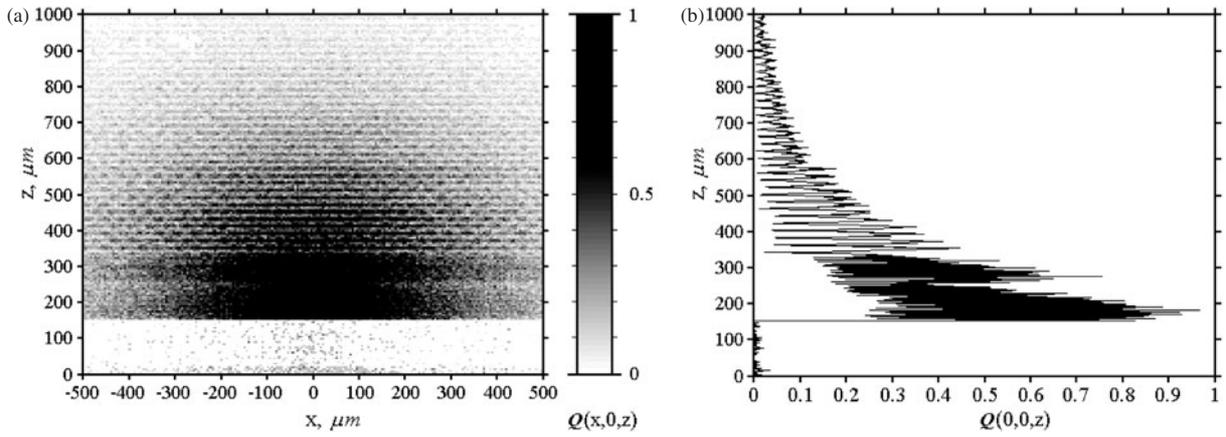

**Figure 7.** Spatial distribution of auto-fluorescence excitation within human skin: (*a*) two-dimensional *x–z* cross-section distribution; (*b*) the distribution profile along the axis *z*. Optical parameters of skin layers are presented in table 1. In the dermal layers the fluorophore distribution closely follows the distribution of collagen bundles modelled by equation (3). The diameters of collagen bundles were chosen as 3, 6, 20, 30 $\mu$m in the PD, upper blood net dermis, dermis and deep blood net dermis, respectively. The diameter of optical source fibre is 200 $\mu$m.

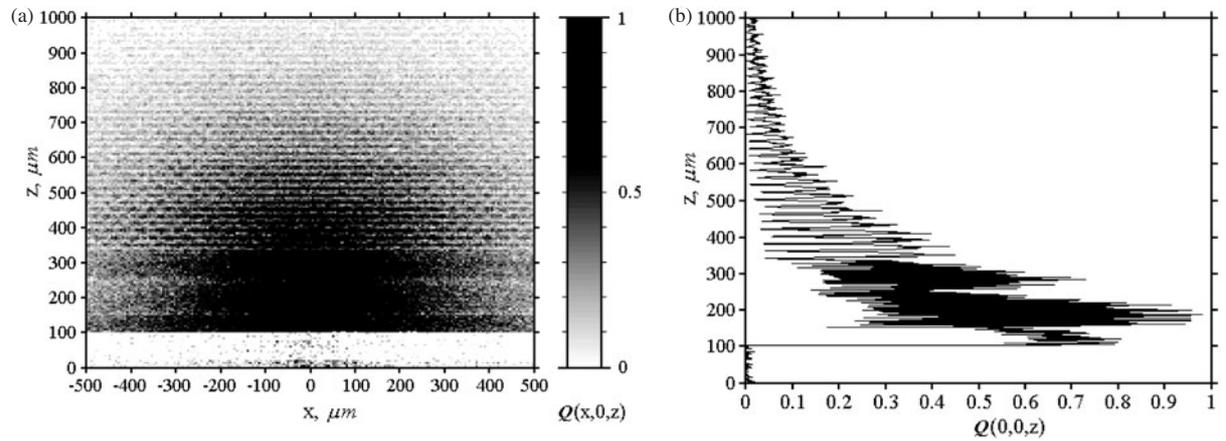

**Figure 8.** Spatial distribution of the fluorescence excitation of the 'tattoo' sensor layer embedded in the epidermis and auto-fluorescence excitation in human skin: (*a*) two-dimensional *x–z* cross-section distribution; (*b*) the distribution profile along the axis *z*. Optical parameters of the model were chosen for 488 nm (see table 1). The diameters of collagen bundles were chosen as 3 $\mu$m, 6 $\mu$m, 20 $\mu$m, 30 $\mu$m in PD, upper blood net dermis, dermis and deep blood net dermis, respectively. The diameter of optical source fibre is 200 $\mu$m.

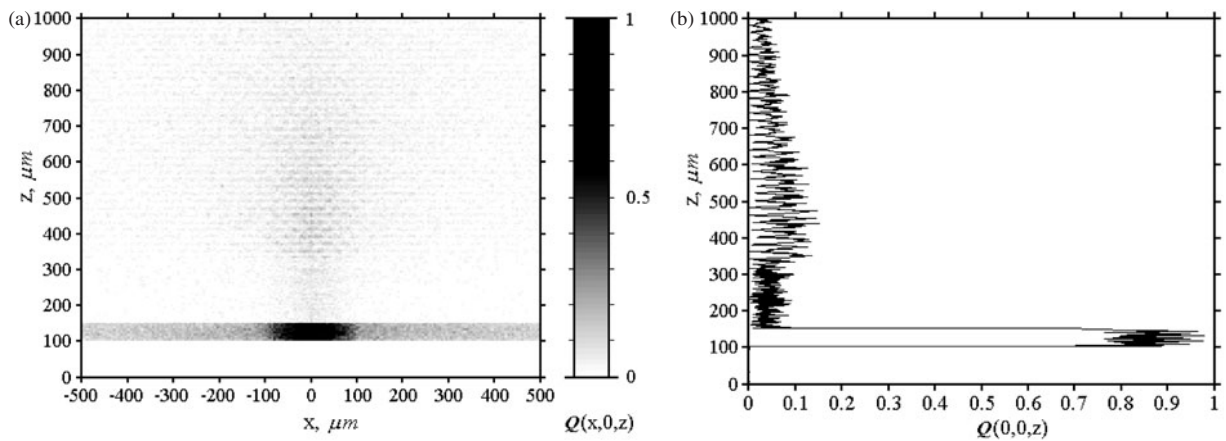

**Figure 9.** Spatial distribution of the fluorescence excitation of the 'tattoo' sensor layer and auto-fluorescence excitation in human skin in the near-infrared spectral region: (*a*) two-dimensional *x–z* cross-section distribution; (*b*) the distribution profile along the axis *z*. The modelled skin tissues optical properties were chosen for 700 nm (see table 1). The absorption coefficients $\mu_a$ of skin layers are taken to be a factor of ten less [8]. Due to the monotonically decreasing scattering of skin tissue with wavelength in the range 450–1100 nm [41–43] the scattering coefficients $\mu_s$ are reduced by a factor of 2–3 (see table 1). The rest of the optical properties and fluorescence parameters are assumed to be independent of wavelength, except for the optical parameters of the sensor layer.

## 4. Conclusions

The novel MC technique for modelling the fluorescence within the human skin has been developed. We demonstrate that the model is able to predict the spatial distribution of the fluorescence/auto-fluorescence excitation within skin. The computational model of skin takes into account spatial distribution of fluorophores that mimics the collagen fibre packing within the dermis. As a result, the observed inhomogeneous porous structure of the fluorescence sources distribution is in a good agreement with the experimental skin fluorescence texture [19]. The results of simulation suggest that the auto-fluorescence background is significantly suppressed in the near-infrared spectral region, whereas the sensor layer fluorescence excitation becomes localized at the adjusted depth (see figure 9). These simulation results are predictable and consistent with the experimental images of the skin tissue auto-fluorescence (see figure 6). In a subsequent work, we will investigate 'tattoo' sampling volume including the effects of tissue scattering, absorption and refractive indices changes.

## Acknowledgments

Authors thank Professor Valery Tuchin (Saratov State University, Saratov, Russia) and Dr Mark Jermy (Cranfield University, Cranfield, UK) for useful discussions during this study. DYC also acknowledges the support of the Department of Optical and Automotive Engineering, School of Engineering, Cranfield University.

## References

[1] Sinichkin Y P, Kollias N, Zonios G I, Utz S R and Tuchin V V 2002 Reflectance and fluorescence spectroscopy of human skin *in vivo Handbook of Optical Biomedical Diagnostics* PM107, ed V V Tuchin (Washington: SPIE Press) pp 727–85
[2] Boas D A, Brooks D H, Miller E L, DiMarzio C A, Kilmer M, Gaudette R J and Zhang Q 2001 Imaging the body with diffuse optical tomography *IEEE Signal Proc. Mag.* **18** 57–75
[3] Beuthan J, Minet O and Muller G 1996 Quantitative optical biopsy of liver tissue *ex vivo IEEE J. Sel. Top. Quant.* **2** 906–13
[4] Richards-Kortum R and Sevick-Muraca E 1996 Quantitative optical spectroscopy for tissue diagnostics *Annu. Rev. Phys. Chem.* **47** 555–606
[5] Bigio I J and Mourant J R 1997 Ultraviolet and visible spectroscopies for tissue diagnostics: fluorescence spectroscopy and elastic-scattering spectroscopy *Phys. Med. Biol.* **42** 803–14
[6] Meglinski I V, Piletsky S A, Greenhalgh D A and Turner A P F 2002 Vanishing 'tattoo' sensors for medical diagnostics *Proc. 2nd International Workshop on Molecularly Imprinted Polymers* (France: La Grande Motte) p 55
[7] Lakowicz J R 1999 *Principles of Fluorescence Spectroscopy* (New York: Plenum)
[8] Tuchin V 2000 *Tissue Optics: Light Scattering Methods and Instruments for Medical Diagnosis* TT38 (Washington: SPIE Press)
[9] Panou-Diamandi O, Uzunoglu N K, Zacharakis G, Filippidis G, Papazoglou T and Koutsouris D 1998 One-layer tissue fluorescence model based on the electromagnetic theory *J. Electromagnet. Wave* **12** 1101–21
[10] Durkin A J, Jaikumar S, Ramanujam N and Richards-Kortum R 1994 Relation between fluorescence-spectra of dilute and turbid samples *Appl. Opt.* **33** 414–23
[11] Patterson M S and Pogue B W 1994 Mathematical model for time-resolved and frequency-domain fluorescence spectroscopy in biological tissues *Appl. Opt.* **33** 1963–74
[12] Nair M S, Ghosh N, Raju N S and Pradhan A 2002 Determination of optical parameters of human breast tissue from spatially resolved fluorescence: a diffusion theory model *Appl. Opt.* **41** 4025–35
[13] Wu J, Feld M S and Rava R P 1993 Analytical model for extracting intrinsic fluorescence in turbid medium *Appl. Opt.* **32** 3585–95
[14] Gandjbakhche A H, Bonner R F, Nossal R and Weiss G H 1997 Effects of multiple-passage probabilities on fluorescent signal from biological media *Appl. Opt.* **36** 4613–19
[15] Muller M G, Georgakoudi I, Zhang Q, Wu J and Feld M S 2001 Intrinsic fluorescence spectroscopy in turbid media: disentangling effects of scattering and absorption *Appl. Opt.* **40** 4633–46
[16] Richards-Kortum R 1995 Fluorescence spectroscopy of turbid media *Optical-Thermal Response of Laser Irradiate Tissue* ed A J Welch and M J C van Gemert (New York: Plenum) pp 667–707
[17] Qu J, MacAulay C, Lam S and Palcic B 1995 Laser-induced fluorescence spectroscopy at endoscopy: tissue optics, Monte Carlo modelling, and *in vivo* measurements *Opt. Eng.* **34** 3334–43
[18] Crilly R J, Cheong W F, Wilson B and Spears J R 1997 Forward-adjoin fluorescence model: Monte Carlo integration and experimental validation *Appl. Opt.* **36** 6513–19
[19] Zeng H, MacAulay C, McLean D I and Palcic B 1997 Reconstruction of *in vivo* skin autofluorescence spectrum from microscopic properties by Monte Carlo simulation *J. Photochem. Photobiol.* B **38** 234–40
[20] Welch A J, Gardner C, Richards-Kortum R, Chan E, Criswell G, Pfefer J and Warren S 1997 Propagation of fluorescent light *Lasers Surg. Med.* **21** 166–78
[21] McShane M J, Rastegar S, Pishko M and Cote G L 2000 Monte Carlo modeling of implantable fluorescent analyte sensors *IEEE Trans. Bio-Med. Eng.* **47** 624–32
[22] Pouge B and Burke G 1998 Fiber-optic bundle design for quantitative fluorescence measurement from tissue *Appl. Opt.* **37** 7429–36
[23] Vishwanath K, Pouge B and Mycek M A 2002 Quantitative fluorescence spectroscopy in turbid media: comparison of theoretical, experimental and computational methods *Phys. Med. Biol.* **47** 3387–405
[24] Yaroslavsky I V and Tuchin V V 1992 Light transport in multilayed scattering media. Monte Carlo modelling *Opt. Spectrosc.* **72** 934–9
[25] Graaf R, Koelink M H, de Mul F F M, Zijlstra W G, Dassel A C M and Aarnoudse J G 1993 Condensed Monte Carlo simulations for the description of light transport *Appl. Opt.* **32** 426–34
[26] Wang L, Jacques S L and Zheng L 1995 MCML—Monte Carlo modelling of light transport in multi-layered tissues *Comput. Meth. Prog. Biol.* **47** 131–46
[27] Keijzer M, Jacques S L, Prahl S A and Welch A J 1989 Light distribution in artery tissue: Monte Carlo simulation for finite-diameter laser beams *Lasers Surg. Med.* **9** 148–54
[28] Boas D A, Culver J P, Stott J J and Dunn A K 2002 Three dimensional Monte Carlo code for photon migration through complex heterogeneous media including the adult human head *Opt. Express* **10** 159–70
[29] Meglinsky I V and Matcher S J 2001 Modelling the sampling volume for skin blood oxygenation measurements *Med. Biol. Eng. Comp.* **39** 44–50

[30] Churmakov D Y, Meglinski I V and Greenhalgh D A 2002 Influence of refractive index matching on the photon diffuse reflectance *Phys. Med. Biol.* **47** 4271–85

[31] Henyey L G and Greenstein J L 1941 Diffuse radiation in the galaxy *Astrophys. J.* **93** 70–83

[32] Meglinski I V, Bashkatov A N, Genina E A, Churmakov D Y and Tuchin V V 2003 The enhancement of confocal images of tissues at bulk optical immersion *Laser Phys.* **13** 65–9

[33] Meglinski I V and Matcher S J 2002 Quantitative assessment of skin layers absorption and skin reflectance spectra simulation in visible and near-infrared spectral region *Physiol. Meas.* **23** 741–53

[34] Sobol' I M 1974 *The Monte Carlo Method* (Chicago and London: The University of Chicago Press)

[35] Zeng H, MacAulay C, Palcic B, McLean D I and Palcic B 1995 Spectroscopic and microscopic characteristics of human skin autofluorescence emission *Photochem. Photobiol.* **61** 639–45

[36] Young A R 1997 Chromophores in human skin *Phys. Med. Biol.* **42** 789–802

[37] Smith L T, Holbrook K A and Byers P H 1982 Structure of dermal matrix during development and in the adult *J. Invest. Dermatol.* **79** 930–1040

[38] Odland G F 1991 Structure of the skin *Physiology, Biochemistry, and Molecular Biology of the Skin* vol 1, ed L A Goldsmith (Oxford: Oxford University Press) pp 3–62

[39] Montagna W, Kligman A M and Carlisle K S 1992 *Atlas of Normal Human Skin* (New York: Springer)

[40] Schneckenburger H, Stock K, Steiner R, Strauss W and Sailer R 2002 Fluorescence technologies in biomedical diagnostics *Handbook of Optical Biomedical Diagnostics* PM107, ed V V Tuchin (Washington: SPIE Press) pp 825–74

[41] Marchesini R, Clemente C, Pignoli E and Brambilla M 1992 Optical properties of *in vivo* epidermis and their possible relationship with optical properties of *in vivo* skin *J. Photochem. Photobiol. B: Biol.* **16** 127–40

[42] Simpson C R, Kohl M, Essenpreis M and Cope M 1998 Near-infrared optical properties of *ex vivo* human skin and subcutaneous tissues measured using the Monte Carlo inversion technique *Phys. Med. Biol.* **43** 2465–78

[43] Doornbos R M P, Lang R, Aalders M C, Cross F M and Sterenborg H J C M 1999 The determination of *in vivo* human tissue optical properties and absolute chromophore concentrations using spatially resolved steady-state diffuse reflectance spectroscopy *Phys. Med. Biol.* **44** 967–81